# Cross-modal topology decodes battery faults from sparse voltage snapshots

Jinwen Li[1,2,3], Yunhong Che[4], Simona Onori[5], Weihan Li[2,3,*], Xiaosong Hu[1,*]

[1]College of Mechanical and Vehicle Engineering, Chongqing University, Chongqing 400044, China

[2]Center for Ageing, Reliability and Lifetime Prediction of Electrochemical and Power Electronics Systems (CARL), RWTH Aachen University, Campus-Boulevard 89, Aachen, 52074, Germany

[3]Institute for Power Electronics and Electrical Drives (ISEA), RWTH Aachen University, Campus-Boulevard 89, Aachen, 52074, Germany

[4]Department of Energy Technology, Aalborg University, Aalborg, 9220, Denmark

[5]Department of Energy Science and Engineering, Stanford University, Stanford, CA 94305, USA

*Corresponding authors: Weihan Li (Weihan.Li@isea.rwth-aachen.de), Xiaosong Hu (xiaosonghu@ieee.org)

## Abstract

Battery safety remains the primary bottleneck for mass electric vehicle (EV) adoption, yet field monitoring is hamstrung by a fundamental asymmetry: complex electrochemical faults must be diagnosed via sparse, low-frequency voltage measurements. Existing methods struggle to resolve the signal ambiguity between overlapping fault modes without hardware upgrades. Here, we demonstrate that these distinct fault fingerprints are not lost, but topologically folded within voltage snapshots. We introduce DeFault, a cross-modal diagnostic framework that mathematically unfolds one-dimensional voltage sequences into multi-dimensional phase-space topologies. DeFault employs a bidirectional cross-attention mechanism that acts as an autonomous, physics-aligned filter, explicitly decoding compounded fault modes that remain fundamentally invisible to sequence-based methods. Validated on a field dataset of 16.4 million data records from 99 in-service EVs, our method achieves an average accuracy of 0.96 and an F1 score of 0.84 for four fault types using only 500-second snapshots (spanning <100 mV). This work proves that high-fidelity, interpretable electrochemical diagnosis is achievable on legacy fleets without new sensors, providing a scalable solution for the battery safety crisis.

## Introduction

The global transition to electrified transportation is accelerating, driven by the dual imperatives of sustainable development and net-zero emissions [1-3]. While lithium-ion batteries have achieved remarkable strides in energy density and cost reduction, facilitating the mass deployment of electric vehicles (EVs), battery safety remains the “Achilles' heel” of the industry [4-6]. Recurring safety incidents precipitated by battery faults, which often evolve silently before triggering catastrophic thermal runaway, have severely eroded consumer confidence and posed liability challenges for manufacturers [7-9]. Such faults inevitably stem from manufacturing imperfections, mechanical abuse, and accumulated stress during prolonged operation [10,11]. Compounded by stochastic user behaviors and heterogeneous evolutions in real-world scenarios, batteries suffer from diverse and complex fault modes [12,13]. However, these physically distinct and opaque mechanisms invariably manifest as highly similar, phenomenologically overlapping fluctuations in standard monitoring signals, rendering accurate early-stage diagnosis one of the most intractable challenges in battery management [14-16].

To untangle this signal ambiguity, conventional diagnosis heavily relies on high-fidelity, multi-dimensional data analysis (e.g., cross-topology voltage, cell-level temperature) [17-19]. However, in the pursuit of cost competitiveness, commercial battery management systems (BMSs) are stripped of the rich sensor suites available in research labs [20,21]. Instead, they rely predominantly on voltage measurements sampled at low frequencies [22,23]. While current and temperature readings are available, these signals typically offer pack-level aggregation or sparse sampling, failing to capture cell-specific state [24,25]. Moreover, the active intervention of the thermal management system often masks subtle thermal anomalies, thereby severely desensitizing temperature signals to early-stage faults [26]. This creates a fundamental disconnect: researchers develop sophisticated diagnostic algorithms assuming abundant signals, while field engineers struggle with sensor scarcity, where sparse, one-dimensional (1D) voltage signals often fail to capture the subtle, multi-dimensional fingerprints of various or evolving faults [27,28]. As a result, unless augmented by laborious, expert-driven manual analysis, conventional methods—ranging from physics-based state observers to advanced deep neural networks—frequently falter in granular fault classification [29-31]. Because hardware upgrades are economically inapplicable to the millions of EVs already deployed, breaking this deadlock requires a computational paradigm shift: we must extract high-dimensional physical insights from the constrained signals we already possess, rather than relying on new sensors.

Circumventing this severe observational limit necessitates a computational paradigm shift: mapping 1D time-series into two-dimensional (2D) topological spaces and leveraging multi-modal artificial intelligence (AI) to decode latent physical structures [32-34]. Inspired by multi-modal human cognition—which integrates diverse sensory representations to construct a robust understanding of reality—this cross-modal perspective offers a pathway to decode complex physical phenomena from sparse data [35-37]. Pioneering efforts have demonstrated that converting voltages into visual formats can indeed expose latent aging and fault patterns to advanced computer vision models [38-40]. However, these emerging single-stream paradigms typically treat visual encoding as a complete replacement for time-series data, capturing structural patterns but discarding the inherent temporal precision of the original sequence. They fail to exploit the collaborative potential of proper cross-modal integration [41-43]. Crucially, existing research remains largely "black box", lacking a transparent link between abstract latent features and the underlying electrochemical physics. Consequently, three critical barriers hinder current progress: the inability of existing feature spaces to disentangle heterogeneous fault fingerprints, the absence of an architecture that actively synthesizes temporal precision with visual topology, and a severe deficit of multi-fault labeled real-world data for validation.

Here, we introduce DeFault, a cross-modal diagnostic framework explicitly designed to achieve high-precision multi-fault diagnosis using only sparse voltage snapshots. By mathematically unfolding 1D voltage residuals into multi-dimensional phase-space topologies—utilizing Gramian Angular Fields (GAF), Markov Transition Fields (MTF), and Recurrence Plots (RP)—we unmask latent nonlinear dynamics and state transitions that are invisible to conventional sequence-only methods. Our DeFault framework integrates these visual textures with temporal trends via a dual-stream architecture with bidirectional cross-attention, enabling a synergistic information gain that resolves the signal ambiguities inherent in legacy sensor data. Specifically, by targeting the universal transition from charging to post-charging, our approach is inherently history-independent, eliminating the cold-start barrier for new or fragmented EV fleets. To bridge the community-wide validation gap, we release a field dataset comprising 47,968 labeled samples from 99 in-service EVs, covering four difficult-to-collect fault types: excessive inconsistency (EI), abnormal self-discharge (ASD), abnormal capacity degradation (ACD), and internal short circuit (ISC). Experimental results demonstrate that our model achieves an average accuracy of 0.96 and an F1 score of 0.84 using only 500-second snapshots (spanning < 100 mV), significantly outperforming state-of-the-art baselines without requiring proprietary historical logs or hardware retrofits. By overcoming the twin challenges of data scarcity and signal entanglement, this work offers a scalable, mechanistically transparent solution for proactive battery guardianship in real-world energy systems.

## Results

### Real-world sensing limitations hinder battery fault diagnosis

In real-world scenarios, battery fault diagnosis is fundamentally bottlenecked by the extreme sparsity of onboard sensor data. To probe the computational limits of diagnosis under these observational constraints, we analyzed an industrial-scale field dataset comprising 16.4 million operational records from 99 in-service EVs (Fig. 1A and 1B). Sampled at a low frequency of 0.1 Hz, this dataset captures realistic charging, discharging, and resting dynamics under highly stochastic user behaviors (Fig. 1C). Crucially, the ground-truth labels for this dataset—encompassing four complex, difficult-to-collect fault modes: EI, ASD, ACD, and ISC (Fig. 1D, Fig. S1 and Table S1)—were determined by engineers after recalling EVs with fault logs and subsequently analyzing, testing, or even disassembling their battery packs (data acquisition and labeling protocols are summarized in Note S1).

This scale of rigorous, real-world validation exposes a severe diagnostic crisis when relying on conventional approaches. Due to strict cost constraints, cell-level voltage is often the sole independent proxy for the internal state of individual cells. However, different fault modes often exhibit overlapping signatures that are indistinguishable in the time domain. As demonstrated by the mean and standard deviation (SD) distributions of cell voltages (Figs. 1E and 1F), the statistical distributions of the categories are similar and overlapping, and it is impossible to identify faults based on typical statistical analysis, let alone distinguish different fault types. Although EVs with ISC show slight deviations in the distribution of voltage standard deviations, which primarily arise from cells affected in the later stages of ISC evolution, most voltage variations remain indistinguishable from those of FF or other faults. This diagnostic ambiguity is further exacerbated by concurrent faults, where a single vehicle exhibits multiple fault types simultaneously, reflecting the uncertain evolution of degradation over time (Fig. S2).

Consequently, conventional 1D time-domain analysis hits a definitive performance ceiling, struggling to disentangle compounded signals where critical diagnostic information is obscured by dimension collapse. Strikingly, however, the distinct textural and structural differences in these voltage curves, while evading basic statistical detection, often remain perceptible to expert visual inspection (Fig. S2). This observation yields a critical insight: the multi-dimensional electrochemical fingerprints of these faults are not permanently lost during low-frequency sampling; rather, they are structurally encoded within the geometric trajectories and temporal textures of the voltage signals. Therefore, instead of exhausting computational resources on increasingly intricate 1D feature engineering, breaking this diagnostic deadlock requires a paradigm shift. We must transition towards a cross-modal topology approach designed to mathematically unfold these hidden dynamics and synergistically fuse sequence fidelity with visual discriminability.

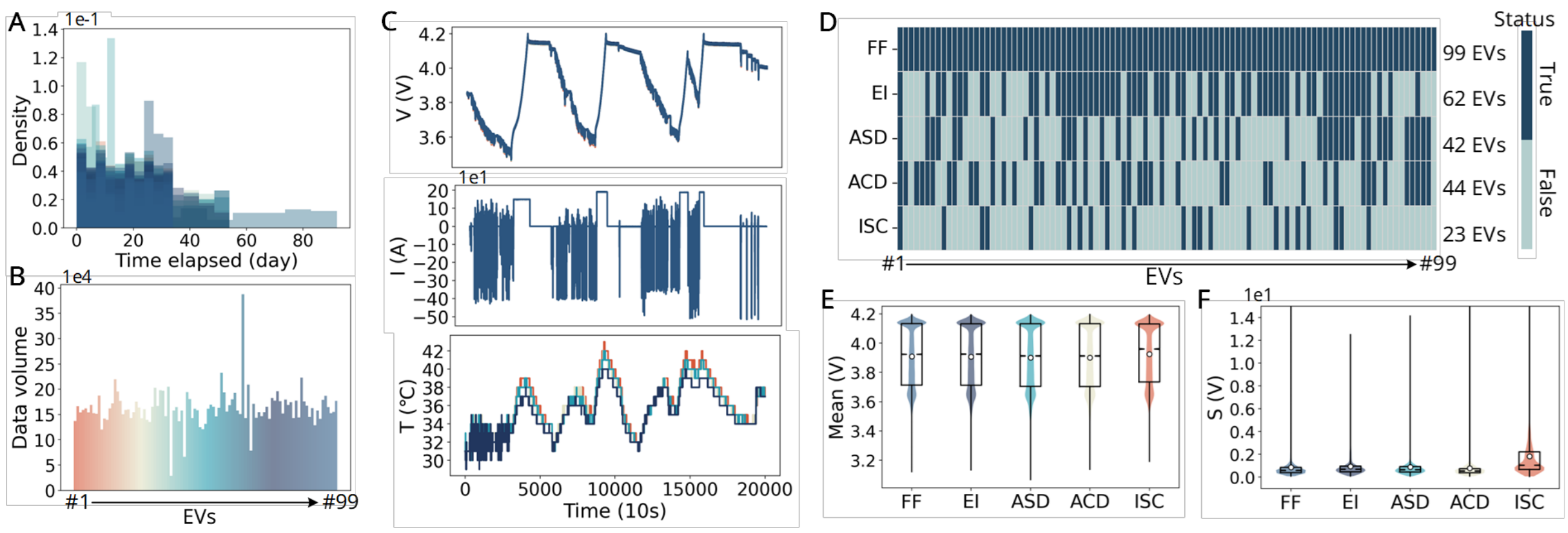

Fig. 1. Real-world data analysis. (A) Elapsed time distribution of collected data. (B) Data volume statistics for each EV. (C) Partial operating data of an EV. From top to bottom, V, I, and T show the changes in voltage, current, and temperature, respectively. The voltage and temperature curves of different colors represent the data from different sensors. (D) Display of EV battery fault types. In the battery packs of most EVs, there are different types of faulty cells. (E) and (F) are the mean and SD distributions of the cell voltages of EVs with different fault types.

**Unlocking latent fault signatures via DeFault framework**

To decipher the complex fault mechanisms obscured within sparse in-service data, we introduce DeFault, a representation-level cross-modal computational framework inspired by human multi-modal cognition (which integrates visual spatial structures and temporal sequences to comprehend complex environments). Mirroring this cognitive capability, DeFault framework redefines the 1D voltage sequence not merely as a scalar timeline, but as a collapsed projection of a multi-dimensional physical state (Fig. 2A). Rather than relying on extensive historical baselines, which are often unavailable for legacy or fragmented EV fleets, we isolate a universal, information-dense electrochemical event: the transition from charging to post-charging. This short-term voltage snapshot acts as a natural dynamic excitation window, powerfully amplifying the distinct depolarization kinetics and thermodynamic signatures of different fault modes. By standardizing input data around these universally accessible events, our method ensures seamless deployment across diverse user habits and new fleets, eliminating the dependency on long-term historical data required by conventional methods.

To disentangle these compounded physical signals, we introduce a sequence-to-image topological pipeline that mathematically unfolds 1D voltage trajectories into 2D visual manifolds (Fig. 2B). Recognizing that raw time-series data lacks the structural diversity necessary for fine-grained multi-fault classification, we map the voltage residuals into a three-channel visual representation (see Methods). By integrating GAF to capture temporal correlations, MTF to profile the transition probabilities of amplitude jumps, and RP to map phase-space chaos, we translate invisible electrochemical variations into unique geometric textures [44-46]. Representative multi-modal feature extraction results (Fig. S3) demonstrate that this topological unfolding effectively exposes fault-specific characteristics, regardless of voltage range or post-charging behavior (resting or driving). Crucially, our feature extraction translates subtle electrochemical variations into distinguishable structures, for example, differentiating the monotonic drift of side-reaction-driven (ASD) from the continuous decay of electronic bypasses (ISC) and the step-down polarization of capacity degradation (ACD). Furthermore, correlation and mutual information analyses confirm that the three image representations are complementary rather than redundant (see Note S3).

To synthesize these multi-modal insights, we propose a dual-stream architecture underpinned by a bidirectional cross-attention mechanism (Fig. 2C). Moving beyond simplistic feature concatenation, DeFault framework establishes a dynamic dialogue between modalities. While specialized convolutional pathways independently preserve the temporal fidelity of the 1D sequence and the spatial topology of the 2D manifolds, the bidirectional cross-attention module actively synthesizes these latent variables. This mechanism allows the model to dynamically calibrate structural visual patterns against local temporal fluctuations, which effectively mimicking the human brain's ability to cross-reference multiple sensory inputs to resolve environmental ambiguity. By mitigating the inherent limitations of single-modality dependence, this bidirectional interaction fundamentally enhances diagnostic precision and sensitivity to early-stage, subtle fault anomalies under highly stochastic real-world operating conditions.

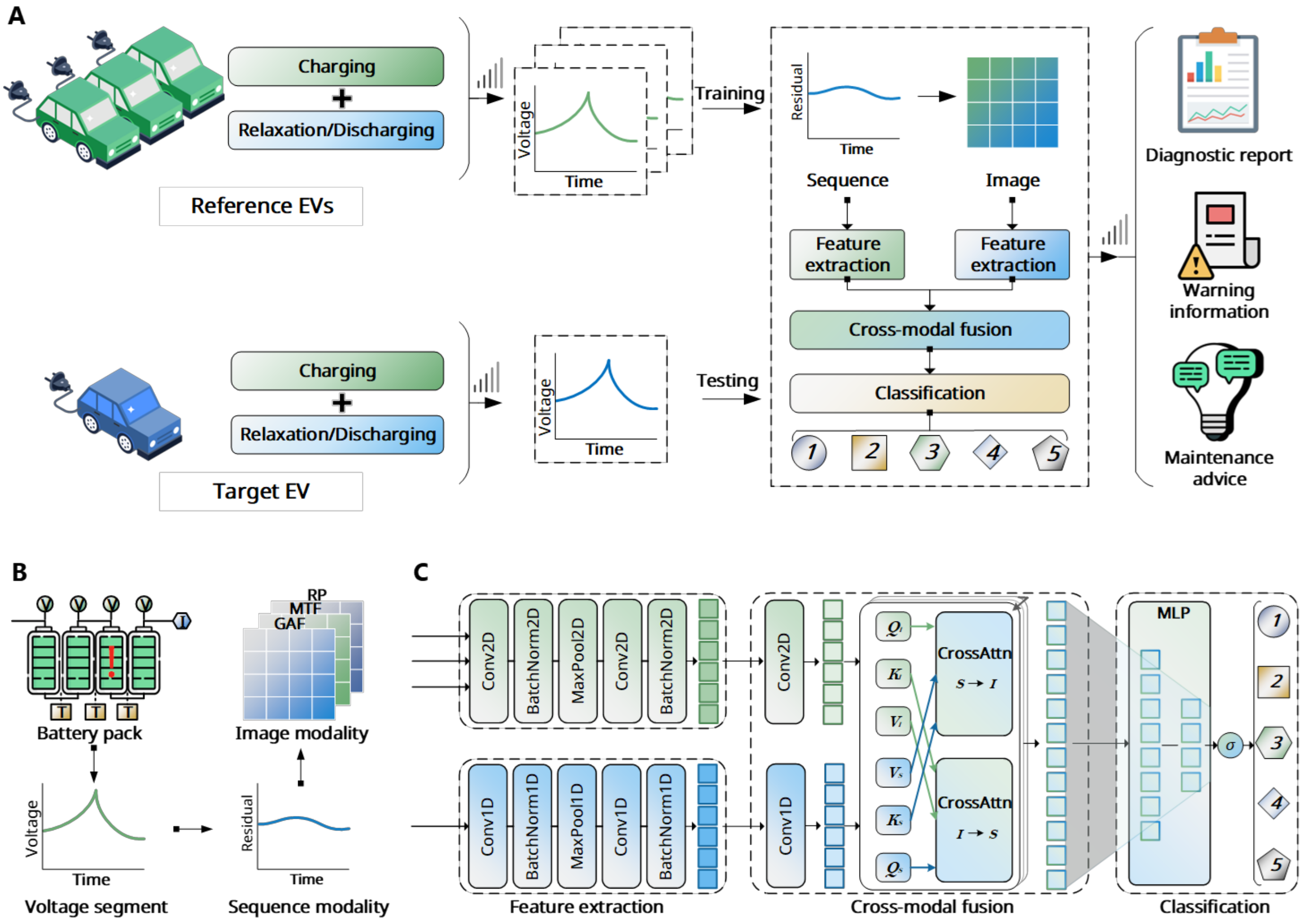

Fig. 2. Illustration of the proposed DeFault framework. (A) Flowchart from data acquisition to fault diagnosis. After EV charging, the voltage snapshot with working condition switching is selected as the input for the cross-modal diagnostic model. The input voltage data undergo multi-modal feature extraction and cross-modal information fusion in the diagnostic

model, and the fault type is finally predicted by the classifier. Based on the results, the diagnostic report, warning information and maintenance advice for the EV battery pack can be generated. The target EV can use the trained model based on the data from reference EVs to diagnose, without its historical data. (B) Construction of multi-modal feature set. The extracted voltage segments are first converted into sequence modes through residual calculation, and then transformed into three-dimensional images based on MTF, GAF and RP. (C) Schematic of cross-modal diagnostic model. The DeFault diagnostic model employs specialized 1D-CNNs and 2D-CNNs to independently extract representations to preserve the unique structure of each modality. Moreover, a bidirectional cross-attention module is employed to enable dynamic interaction, adjustment, and fusion of latent variables across modalities. Furthermore, the fault type is predicted by using a multi-layer perceptron (MLP) classifier, with the highest-probability class serving as the diagnostic result.

**Multi-fault diagnostic using short-term voltage snapshots**

One of central challenge in deploying computational diagnostics in real-world physical systems is determining the minimal observational window required to capture a complete mechanistic fingerprint. To probe this computational limit, we evaluated DeFault framework across constrained input segments ranging from 100 s to 1000 s. All segments are rigidly aligned to the end of charging, ensuring that the critical transition into the post-charging relaxation phase is processed under a unified thermodynamic baseline. As detailed in Fig. 3A–C, the framework demonstrates exceptional discriminative robustness under extreme data sparsity (the evaluation metrics are detailed in Note S4). Excluding the absolute minimum 100 s baseline, the average multi-class accuracy across all temporal settings stably exceeds 0.960. Furthermore, the model maintains a true positive rate (TPR) above 0.95 while bounding the false positive rate (FPR) within 0–0.03. This confirms that even under extreme temporal compression (e.g., merely 10 sampling points in the 100 s setting), the multi-modal topological projection successfully preserves high diagnostic sensitivity while keeping false positives low.

Crucially, our computational analysis reveals a fundamental information saturation threshold at precisely 500 seconds (Fig. 3D and 3E). Extending the observational window beyond this temporal boundary yields rapidly diminishing diagnostic returns: expanding the snapshot to 1000 s quadruples the computational overhead (a 4.82× increase in training time) but yields a marginal accuracy gain of only 0.73%. This algorithmic plateau exposes a profound physical reality: the essential electrochemical dynamics—specifically, the discharge of the double-layer capacitor and the evolution of initial concentration polarization—are completely topologically encoded within just a few minutes of relaxation. By identifying 500 s as the optimal computational-physical nexus, DeFault framework guarantees diagnostic confidence (>92.5% identification rate across all fault categories) while minimizing data transmission overhead, ensuring rapid edge-computing responsiveness for field deployment. Furthermore, when compressed into a binary fault detection gatekeeper, the model maintains robust generalized performance, validating its utility as a primary safety perimeter (Fig. S5).

From a system safety perspective, the model's performance on ISC is paramount, as these are the primary precursors to thermal runaway. Remarkably, even at the shortest input lengths, the cross-modal architecture consistently identifies >96.2% of ISC samples. This high-fidelity detection indicates that the chaotic phase-space disruptions caused by micro-shorts create highly distinctive topological textures, enabling the attention mechanism to inherently prioritize safety-critical anomalies. Beyond static classification, the framework functions as a quantitative, high-resolution risk monitor. As illustrated in Fig. 4F, the model's probabilistic outputs accurately track the continuous, subtle degradation trajectory of a cell transitioning from early EI to fully developed ASD. Notably, in the highlighted case, the model successfully isolated a specific early-stage fault driven by a residual deviation of merely 3.02 mV. This extraordinary sensitivity confirms that our topology-driven computational approach does not rely on simple thresholding, but rather actively mines complex, nonlinear correlations to detect anomalies that are invisible to conventional BMS logic.

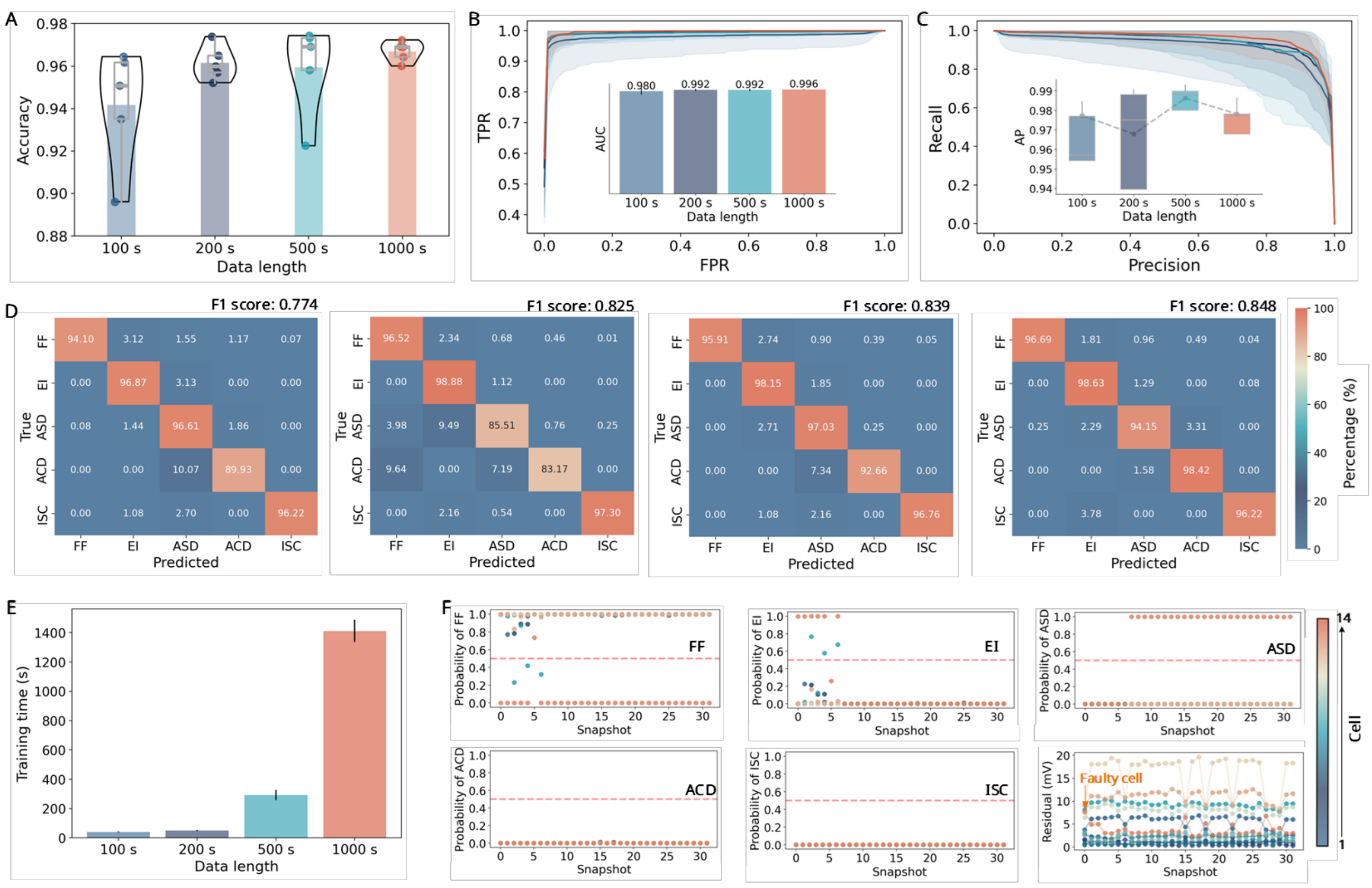

Fig. 3. Evaluation of multi-fault diagnosis with various input data. (A) The colored dots represent the validation accuracy of each 5-fold validation, and the bars represent their average values. (B) Receiver operating characteristic (ROC) curve and average area under the curve (AUC). (C) Precision-Recall (PR) curve and average precision (AP). In the ROC and PR curves, the solid line and shaded interval represent the mean and 95% confidence interval from the 5-fold validation, respectively. (D) The confusion matrix is used to evaluate how well the model can diagnose each fault type, and it shows the average result of 5-fold validation. (E) The training time of the proposed model. (F) Fault diagnosis results based on each snippet of an EV with ASD. It shows the probability that snapshots are classified into different categories (FF, EI, ASD, ACD and ISC). The category with the highest probability is the diagnosis result. In this case, due to unknown factors, the model made three false diagnoses (total sample size: 448; error rate: 0.67%). The subgraph in the lower-right corner showing the residuals shows the average residual voltage difference between each cell and the median voltage.

**Overcoming the diagnostic accuracy paradox via cross-modal topology**

To rigorously evaluate our DeFault framework against existing computational paradigms, we conducted extensive benchmarking across a spectrum of state-of-the-art architectures, including sequence-only, image-only, and conventional multi-modal fusion strategies (early, mid, and late-stage fusion; see Note S5 for implementation details). Evaluated at the critical 500-second information saturation threshold (Fig. 4), this comparative analysis exposes a fundamental vulnerability in conventional diagnostics when applied to highly imbalanced, real-world physical systems.

Specifically, our analysis reveals that conventional unimodal approaches—regardless of their underlying algorithmic complexity—succumb to a severe accuracy paradox. In the highly skewed data environments typical of industrial fleets, these models achieve deceptively high overall accuracy primarily by over-fitting to the abundant fault-free baseline. Consequently, they routinely fail to pinpoint rare, safety-critical anomalies, creating a dangerous false sense of security. This algorithmic blind spot underscores a fundamental physical limitation: relying solely on a single observational modality (either purely temporal or purely visual) imposes a strict information bottleneck, stripping the model of the multidimensional context required to decouple phenomenologically similar but mechanistically distinct faults.

Even among multi-modal baselines, our biologically inspired cross-attention mechanism proves uniquely capable of breaking this bottleneck. While late-stage decision fusion emerged as the strongest conventional competitor (achieving the second-best performance), it essentially functions as a post-hoc voting mechanism, entirely failing to exploit the underlying synergistic correlations between modalities. In stark contrast, DeFault framework establishes a deep, dynamic dialogue. By allowing the network to continuously cross-reference temporal kinetic fluctuations against visual thermodynamic topologies—much like a human expert validating a subtle voltage dip against a broader degradation texture—our approach inherently dynamically corrects signal ambiguities. Quantitatively, this active cross-modal calibration yields a substantial 6.33% improvement in the average F1 score over the best-performing fusion baseline. In the context of battery safety, where a single missed ISC can trigger catastrophic thermal runaway, this drastic reduction in false negatives unequivocally justifies the marginal increase in computational parameters.

Furthermore, the robustness of this topological paradigm is most pronounced under conditions of extreme observational scarcity (Fig. S6). When constrained to a mere 100-second snapshot (just 10 sampling points), our DeFault framework acts as a powerful computational amplifier. It achieves a remarkable performance leap, boosting the average accuracy and F1 score by 61.85% and 32.29%, respectively, compared to state-of-the-art unimodal baselines. This extreme-sparsity resilience confirms that our approach does not merely stack data matrices; rather, it mathematically unfolds hidden nonlinear dynamic structures that remain entirely invisible to conventional sequence models. By leveraging cross-modal cognition to distinguish subtle electrochemical nuances, DeFault framework decisively solves the class imbalance and signal entanglement challenges inherent to data-starved real-world diagnostics.

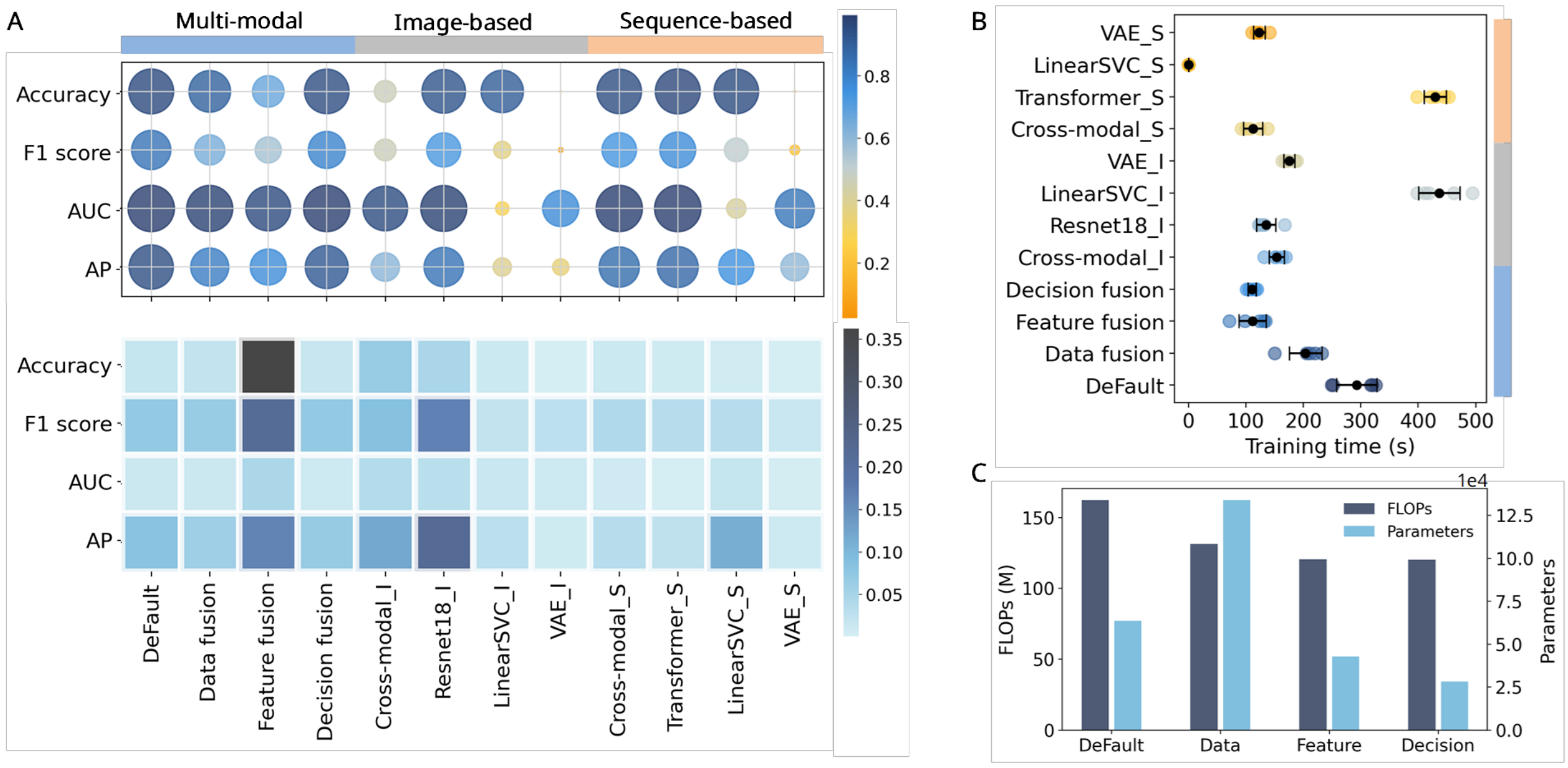

Fig. 4. Performance comparisons of multi-fault diagnosis with different baselines. These validation results are based on voltage segments of length 500 s. In the comparison methods, the multi-modal baselines comprise three mainstream fusion strategies: data fusion (early stage), feature fusion (mid stage), and decision fusion (late stage). The feature fusion baselines serve as ablations of our method. They are mid-stage fusion models that perform shallow feature fusion rather than cross-attention-based fusion. For single-modality baselines, in addition to ablations of our method using sequence-only or image-only variants, we include a classical Variational Autoencoder (VAE) for unsupervised learning and Linear Support Vector Classifier (LinearSVC) for traditional machine learning. We also compare against modality-specific ResNet18 for images and Transformer for sequences. Moreover, the markings I and S indicate that the input data are only image and sequence modality, respectively. Cross-modal representation: the modal fusion structure of this model is the same as that of the proposed method; the difference is that the feature input structures of its two streams are either an image feature module or a sequence feature module. (A) Multi-fault diagnosis performance. The top circular heatmaps and the bottom square heatmaps represent the mean and standard deviation of the performance metrics, respectively. (B) Model training time. (C) Model complexity analysis with multi-modal input models.

**Decoding the physical interpretability of cross-modal topologies**

A persistent criticism of deep learning in safety-critical applications is its black-box nature. To bridge the gap between algorithmic prediction and engineering trust, we interrogated the model's internal decision pathways, visualizing how DeFault decodes abstract computational variables into profound physical insights. Since these latent variables reside in a high-dimensional abstract space, we employed t-distributed Stochastic Neighbor Embedding (t-SNE) to project them into a human-readable 2D manifold [47]. To quantitatively track this spatial evolution across network layers, we introduced a separation score, defined as the ratio of inter-class distance to the sum of intra- and inter-class distances (see Note S6).

The resulting visualization unfolds a striking narrative of high-dimensional physical disentanglement. Mapping the latent variables at the initial unimodal extraction stages (post-CNN backbones) reveals a profound thermodynamic and kinetic entanglement (Fig. 5A and 5B). The distributions of fundamentally different fault types collapse into a nebulous, overlapping statistical cloud rather than forming distinct clusters. This visual dimension collapse explicitly confirms our core hypothesis: shallow, isolated modalities lack the structural dimensionality required to resolve the phenomenological degeneracy of complex battery faults. At this stage, the uncoupled network struggles to separate critical electrochemical signals from stochastic baseline noise, yielding a severely depressed separation score.

A dramatic computational phase transition occurs immediately at the cross-attention interaction layer. Once the bidirectional fusion mechanism is engaged, the latent landscape undergoes a dramatic transformation. As illustrated in Fig. 5 B(e), the previously entangled fault categories crystallize into well-separated islands. Quantitatively, this disentanglement is absolute: the separation score surges by more than 80% post-interaction compared to the isolated unimodal stage (Fig. 5C). This abrupt leap in separability provides physical proof that the cross-modal mechanism is not merely adding features, but actively synthesizing complementary information. By the final classification layer, the separation score improves by an additional 7.06%, further cementing the model's confidence. This visualization confirms that our DeFault framework succeeds not by memorizing data, but by learning fundamental, robust representations of fault physics, making the diagnostic process transparent and interpretable.

Crucially, we demonstrate that this high-dimensional separation is grounded in battery physics rather than statistical artifacts. By mapping the cross-attention weights back to the original operational sequences, we expose exactly how the model's “computational eye” interrogates the physical system (Fig. 5D–H). Moreover, the complete decision-making trajectory from raw voltage snapshots to final diagnostic probabilities is shown in Fig. S7. Astoundingly, without any explicit programming of battery physics, the cross-attention mechanism autonomously acts as a physics-aligned filter. Specifically, for the ACD case shown in Fig. 5D, the attention heads intensely focus on the voltage relaxation phase immediately following charging. This aligns with electrochemical principles, as this phase contains critical information about concentration polarization and ohmic voltage drop, which are direct indicators of active material loss and lithium inventory depletion. Conversely, for kinetically driven faults like ISC, the focus shifts to the continuous voltage decay trajectory driven by the electronic bypass. By synergizing these specific temporal dynamics with global visual topological representations, the model generates robust, clear decision boundaries. This confirms that DeFault transcends “black box” mapping, effectively discovering and decoding the phenomenological fingerprints of battery faults.

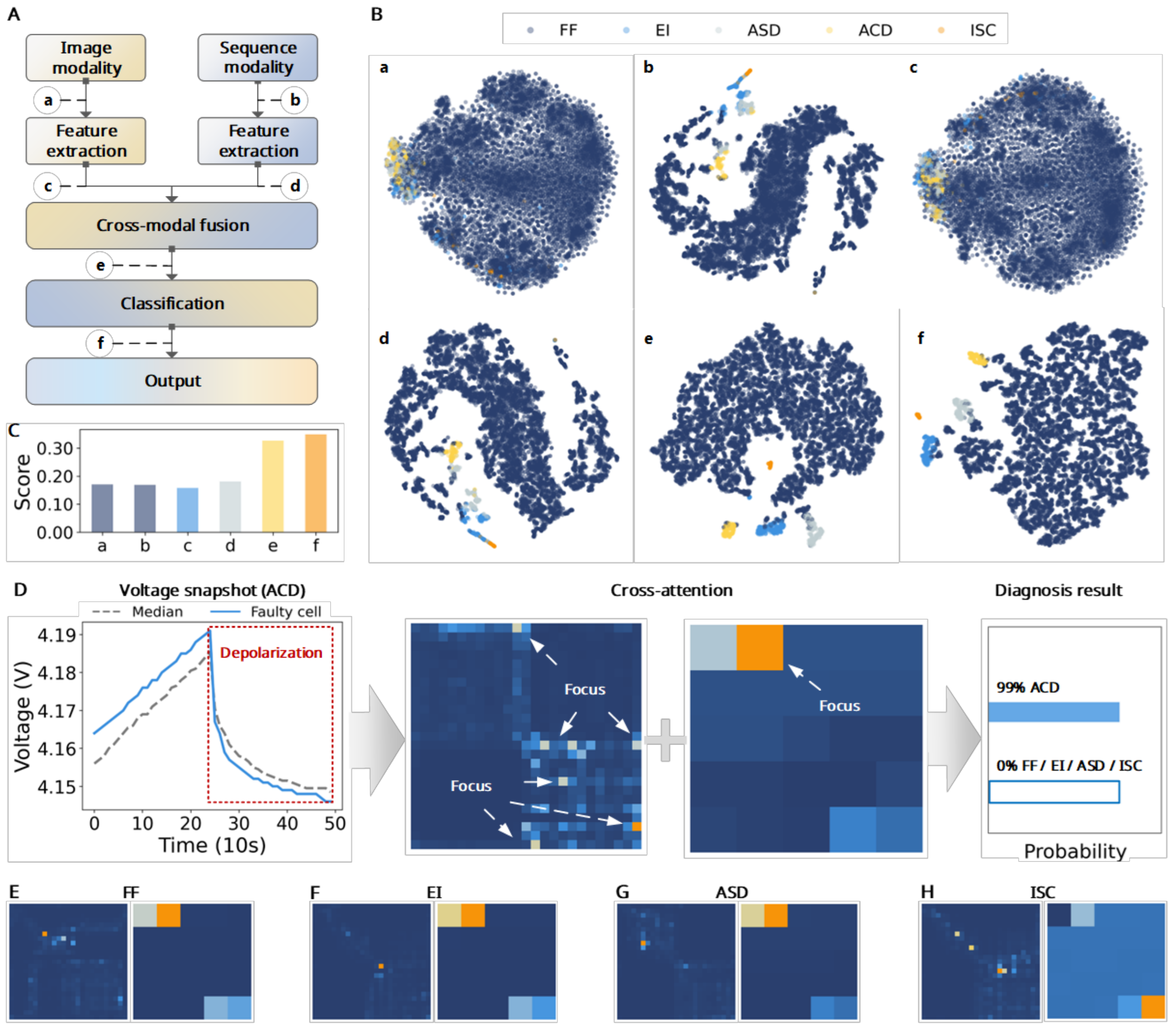

Fig. 5. Visualization of the cross-attention diagnostic pipeline. (A) Schematic of the latent variable extraction points (a–f) within the model architecture. (B) Evolution of feature disentanglement visualized via t-SNE. Points a and b represent unimodal features (Image and Sequence), which show high entanglement. Points c and d represent features after independent extraction but before fusion. Point e (inferred latent variable) reveals a dramatic separation of fault clusters after cross-modal attention, culminating in the final output f. (C) The separation score across different model stages, quantifying the disentanglement capability. (D)-(H) Interpretability analysis using cross-attention maps for five samples of ACD, FF, EI, ASD, and ISC, respectively. The heatmaps illustrate the bidirectional attention weights: the left map represents the visual topology querying temporal details, while the right map represents temporal trends seeking structural confirmation from the image. (D) Detailed analysis of an ACD sample. The high-attention regions (white/yellow blocks) precisely align with the depolarization phase in the voltage snapshot (red box), indicating the model correctly identifies the polarization buildup associated with capacity fade.

## Discussion

The electrification of global transport is currently hamstrung by a technological asymmetry: while battery chemistries become increasingly sophisticated, in-service monitoring remains tethered to low-fidelity, scalar voltage sensors. Our work resolves this paradox by demonstrating that the limitation lies not in the sensor's hardware, but in the dimension of analysis.

By mathematically elevating 1D time-series into 2D visual structures, we unlock latent dynamic electrochemical fingerprints that are folded and invisible in the original sequence. Our DeFault framework actively synthesizes the instantaneous precision of temporal trends with the structural discriminability of visual patterns. Validated on a released field dataset of 99 EVs, the framework achieves an average accuracy of 0.96 and an F1 score of 0.84 using only 500 s voltage snapshots. These results provide definitive proof that low-frequency data contains sufficient diagnostic information, provided it is decoded through the correct high-dimensional cross-modal lens.

Beyond shattering the performance ceiling of sparse diagnostics, our DeFault framework redefines the role of AI in safety-critical physical systems, transitioning it from a “black-box” oracle to a transparent, physics-aware microscope. As detailed in interpretability analysis, the model successfully disentangles kinetically dominant faults (e.g., early-stage ISC) from thermodynamically governed degradation (e.g., ASD). By providing transparent decision pathways that autonomously align with governing electrochemical principles, this approach bridges the critical trust gap between algorithmic predictions and engineering intuition. Furthermore, by achieving high precision with only a short-term voltage snapshot, our approach eliminates the cold-start problem for new or fragmented fleets, enabling scalable, cloud-based supervision without proprietary historical logs.

Equally critical to our methodological innovation is the democratization of high-fidelity data. The severe deficit of publicly available, fault-labeled field data has long bottlenecked advanced model verification in battery informatics. By releasing a rigorously validated field dataset comprising 16.4 million records and four difficult-to-collect fault types from 99 in-service EVs, this work establishes a much-needed ground truth. We anticipate this dataset will serve as a foundational benchmark, accelerating the community-wide development of robust AI-for-Science diagnostics.

Looking forward, while this framework establishes a robust baseline for ternary chemistries, the intrinsic electrochemical properties of Lithium Iron Phosphate (LFP)—notably the flat voltage plateaus—define a unique signal-to-noise frontier. Crucially, our work introduces a transferable topological paradigm. By capturing the fundamental fault dynamics within data-rich fleets, we establish the essential feature representations required for cross-chemistry adaptation. This knowledge architecture provides a clear methodological pathway to bridge the industry-wide LFP labeling deficit through domain adaptation and transfer learning. Ultimately, this cross-modal topological approach serves as a scalable foundation for high-fidelity computational guardianship, extending reliable, interpretable safety supervision to diverse chemistries, data-starved environments, and the broader spectrum of next-generation intelligent energy systems.

## Methods

### Multi-modal feature set

To extract features that are both sensitive to early faults and information-rich from raw voltages, we constructed a novel representation-level multi-modal feature set. For each cell, the raw voltages are referenced to the median voltage across the pack to generate residuals, thereby enhancing the visibility of subtle anomalies, as follows [48]:

$$R_{pack} = [R_1, R_2, \dots, R_n] = [v_1 - \tilde{v}, v_2 - \tilde{v}, \dots, v_n - \tilde{v}], \tag{1}$$

where $v_n$ denotes the voltage of the *n*-th cell, $\tilde{v}$ is the median voltage, and $R_n$ is the residual used to characterize the sequence modality of the *n*-th cell.

For the image modality, three different encoding techniques, GAF, MTF and RP, are used to convert the 1D residual sequence into 2D images to capture complementary perspectives of the residual dynamics. Prior to encoding, the residual sequence of length $t$, $R = [r_1, r_2, \dots, r_t]$, is scaled to the range [-1, 1]:

$$\tilde{r}_t = \frac{(r_t - \max(R)) + (r_t - \min((R)))}{\max(R) - \min(R)}. \tag{2}$$

Then, the GAF matrix is generated by computing the cosine of the angular summation, capturing temporal correlations and smooth local transitions [49]:

$$GAF_{ij} = \cos\left(\arccos(\tilde{r}_i) + \arccos(\tilde{r}_j)\right). \tag{3}$$

MTF describes the dynamic statistical characteristics of sequences by discretizing and calculating the transition probabilities between states [50]. The scaled residual sequence $\tilde{R} = [\tilde{r}_1, \tilde{r}_2, \dots, \tilde{r}_t]$was discretized into $Q$quantile bins. Each residual $\tilde{r}_i$was assigned to a state $s_i \in (q_1, q_2, \dots, q_Q)$. The MTF matrix is then computed from the first-order Markov transition matrix $W$, where each element represents the transition probability along the temporal axis:

$$MTF_{ij} = w_{s_i s_j}, \tag{4}$$

where the element at position $(i, j)$ in the MTF matrix is the probability that the state at time *i* will transition to the state at time *j*.

RP is a binary matrix; it reconstructs the scaled residual sequence $\tilde{R} = [\tilde{r}_1, \tilde{r}_2, \dots, \tilde{r}_t]$ into a phase space of dimension $m$ with time delay $\tau$ $\vec{x}_i = \left[r_i, r_{i+\tau}, \dots, r_{i+(m-1)\tau}\right]^T$. Following this, RP can be expressed as [51]:

$$RP_{ij} = \Theta(\epsilon - \left\|x_i - x_j\right\|, \tag{5}$$

where $\Theta$ is the Heaviside step function and $\epsilon$ is the threshold for recurrence. RP reveals nonlinear dynamical properties such as chaos and stability in sequences.

For each cell voltage sequence, three images (GAF, MTF, RP) are generated synchronously. They provide complementary views of correlation, stochasticity, and recurrence, thereby depicting the complete dynamic behavior of the sequence modality. These three single-channel images are then combined as a three-channel input and fed into a CNN-based feature extractor.

**Cross-modal multi-fault diagnosis model**

To fully exploit the rich information contained in the multi-modal feature set, we propose a dual-stream cross-modal fault diagnosis model. The architecture consists of four key components: a visual feature extractor, a temporal feature extractor, a bidirectional cross-attention fusion module, and a classifier. To project the raw inputs from different modalities into a shared latent feature space, two dedicated convolutional neural networks (CNNs) are employed. Its convolutional-pooling modules are designed to learn spatially local features. Defining a 2D-CNN convolutional-pooling as $l$, the output can be described as follows [52]:

$$y_k^l = f\left(b_k^l + \sum_{i=1}^{C^{l-1}} cov2D\left(W_{k,i}^l, S_i^{l-1}\right)\right), k = 1,2, \dots, M, \tag{6}$$

where $S^{l-1}$ represents the *i*-th input image feature map from the $l-1$ module, $b_k^l$ and $W^l$ represent the bias and weights for the *k*-th output feature map, respectively. $f(\cdot)$ is the activation function, and *M* is the number of output feature maps, determined by the number of kernels applied. For sequential modality, it is handled by a 1D-CNN. The feature extraction process of 1D-CNN and 2D-CNN is similar; the only difference is the use of a 1D convolution kernel and a single-channel structure.

To facilitate feature exploration across different modalities, we utilize the cross-attention between image and sequence modalities in a bidirectional manner. First, the image modality $F_{img}$ and sequence modality $F_{seq}$ are mapped into the he common subspace $\mathcal{Q}_{img}, \mathcal{K}_{img}, \mathcal{V}_{img} \in \mathbb{R}^{N_{img} \times D}$ and $\mathcal{Q}_{seq}, \mathcal{K}_{seq}, \mathcal{V}_{seq} \in \mathbb{R}^{N_{seq} \times D}$ with dimension *D*, respectively, by using pointwise convolution. Then, bidirectional cross-modal fusion is computed via Multi-Head Attention (MHA) [53].

$$\begin{cases} Attn_{img \to seq} = MHA\left(\mathcal{Q}_{img}, \mathcal{K}_{seq}, \mathcal{V}_{seq}\right) = softmax\left(\frac{\mathcal{Q}_{img}(\mathcal{K}_{seq})^T}{\sqrt{D}}\right)\mathcal{V}_{seq} \\ Attn_{seq \to img} = MHA\left(\mathcal{Q}_{seq}, \mathcal{K}_{img}, \mathcal{V}_{img}\right) = softmax\left(\frac{\mathcal{Q}_{seq}(\mathcal{K}_{img})^T}{\sqrt{D}}\right)\mathcal{V}_{img} \end{cases}. \tag{7}$$

Moreover, the outputs are combined with their original inputs through residual connections, followed by layer normalization (LN) and non-linear activation, enhancing the features while preserving stability during training:

$$\begin{cases} \tilde{F}_{img} = LN(\mathcal{Q}_{img} + Attn_{img \to seq}) \\ \tilde{F}_{seq} = LN(\mathcal{Q}_{seq} + Attn_{seq \to img}) \end{cases}. \tag{8}$$

Finally, the enhanced features ($\tilde{F}_{img}$ and $\tilde{F}_{seq}$) are aggregated via global average pooling (GAP) and concatenated to form the fused representation $z = [GAP(\tilde{F}_{img}); GAP(\tilde{F}_{seq})]$.

Furthermore, the fused representation *z* is further refined through a multi-layer fusion processor with dropout regularization, resulting in the prediction of fault class probabilities $\hat{y}$ by the classifier.

**Sample balancing and model training**

To address sample imbalance in real-world scenarios, we adopted a dual sample-balancing strategy to enhance the model's generalization and improve recognition of minority fault classes. On the one hand, a weighted random sampling strategy is applied for the training dataset. The weight assigned to each sample is defined as [54]:

$$weights = \beta \times \frac{1}{C[labels]} + (1 - \beta) \times \frac{1}{len(weights)}, \tag{9}$$

where *C* represents the number of categories, $\beta \in [0,1]$. When $\beta = 1$ represents complete balance, and conversely when $\beta = 0$, it corresponds to the original distribution. $\beta$ is set as 0.5 in this study. On the other hand, class weights are incorporated into the loss function to establish a cost-sensitive learning mechanism. This further alleviates the imbalance problem by penalizing misclassification of minority classes more heavily, thereby improving fault recognition performance across all categories. The proposed model is trained end-to-end by minimizing the cross-entropy loss between predictions and true labels, and its weighted form is defined as [55]:

$$loss = -\frac{1}{B}\sum_{i=1}^{B} \alpha_{L_i} \log\left(\frac{\exp(z_{i,L_i})}{\sum_j^C \exp(z_{i,j})}\right). \tag{10}$$

Here, $L_i$ represents the true label of the *i*-th sample, $z_{i,j}$ represents the logit output of sample *i* on category *j*, and *B* is the batch size. The category weight $\alpha_C$ is calculated through inverse frequency weighting $\alpha_C = N/C \cdot n_C$. This setting ensures that the contribution of each class to the total loss is relatively balanced.

## Data availability

The data used in this work will be made publicly available upon final publication.

## Acknowledgements

X. H. acknowledges the financial support from the Technical Innovation and Application Development Special Program of Chongqing Major Project (CSTB2024TIAD-STX0031), National Natural Science Foundation of China (U23A20327, 52402465), the project of basic research funds for central universities (2023CDJQCZX-001), and the relevant national project. W. L. acknowledges the financial support from the research project "SafeDaBatt" (03EMF0409A), funded by the German Federal Ministry of Transport (BMV). J. L. acknowledges the financial support from the China Scholarship Council (202506050053) during his research stay at RWTH Aachen University. Y. C. acknowledges the financial support from the Novo Nordisk Foundation (NNF24OC0088261).

## Declarations

The authors declare that they have no known competing financial interests or personal relationships that could have appeared to influence the work reported in this paper.

# Supplementary Information:
# Cross-modal topology decodes battery faults from sparse voltage snapshots

Jinwen Li[1,2,3], Yunhong Che[4], Simona Onori[5], Weihan Li[2,3,*], Xiaosong Hu[1,*]

[1]College of Mechanical and Vehicle Engineering, Chongqing University, Chongqing 400044, China

[2]Center for Ageing, Reliability and Lifetime Prediction of Electrochemical and Power Electronics Systems (CARL), RWTH Aachen University, Campus-Boulevard 89, Aachen, 52074, Germany

[3]Institute for Power Electronics and Electrical Drives (ISEA), RWTH Aachen University, Campus-Boulevard 89, Aachen, 52074, Germany

[4]Department of Energy Technology, Aalborg University, Aalborg, 9220, Denmark

[5]Department of Energy Science and Engineering, Stanford University, Stanford, CA 94305, USA

*Corresponding author: Weihan Li, Xiaosong Hu

**Supplementary Note 1: Data acquisition and labeling protocols**

In this study, we analyzed a large-scale field dataset from 99 EVs, recorded by the onboard battery management system (BMS) and uploaded to a cloud platform. All 99 EVs share identical battery pack specifications provided by an anonymous BMS manufacturer. Due to strict commercial confidentiality, specific details regarding vehicle applications, operational regions, and battery electrochemical materials cannot be disclosed. Each pack consists of 14 series-connected ternary cells, equipped with 14 voltage sensors, 4 temperature sensors, and 1 current sensor. Each data record contains cell voltages, probe temperatures, pack current, pack voltage, time, charging status, and the voltage and temperature extrema (maximum and minimum values). In addition, the dataset was preprocessed to address missing values and remove anomalous entries before analysis (see Supplementary Note 2).

The dataset and corresponding fault labels used in this study were curated from a large-scale fleet of 99 in-service electric vehicles. It is important to explicitly state that due to strict commercial confidentiality agreements with the BMS manufacturer, the complete, proprietary diagnostic method cannot be disclosed. Instead, the labels provided in this dataset were retrospectively derived and summarized from engineering accident analysis reports, fault recall logs, and disassembly test records provided by engineers. These labels represent the final conclusion reached by expert engineers after investigating specific safety incidents, regular inspection, or customer complaints.

While the exact proprietary protocols remain confidential, the diagnostic logic summarized from the engineering reports relies on fundamental electrochemical principles. Engineers confirmed these faults by identifying specific violations of voltage consistency and evolution laws during recalls or maintenance (based on engineering experience and disassembly test of some battery packs). The criteria for assigning labels to each fault are roughly as follows:

**Excessive inconsistency (EI):** EVs were recalled by the supervision center of EV due to the continuous triggering of BMS cell imbalance error codes. Such faults were defined by the criterion when the voltage traces exhibit persistent outliers without the persistent leakage characteristic of the ISC.

**Abnormal self-discharge (ASD):** EVs were recalled because of complaints from customers about range drops after parking. Unlike the ohmic leakage of ISC, ASD was identified by reversible voltage drift. Engineers noted that these cells showed accelerated voltage drops at rest after charging (driven by thermodynamic instability/side reactions), distinguishing them from the constant linear decay of micro-shorts.

**Abnormal capacity degradation (ACD):** These EVs exhibited a sudden state of health drop in cloud regulation or a sudden range drop problem in the feedback of users. This fault is identified via the charging high and discharging low phenomenon. The faulty cell hit the upper voltage cutoff first during charging (due to polarization buildup) and the lower cutoff first during discharging or resting, confirming a physical loss of active material or lithium inventory.

**Internal short circuit (ISC):** EVs are recalled following high-risk warnings from cloud monitoring platform or EVs have experienced a thermal runaway incident and are diagnosed by accident traceback. A cell was labeled ISC if it exhibited a continuous, load-independent linear voltage decay that significantly exceeded the pack average, confirmed by forensic reconstruction to be caused by separator failure or dendrite penetration.

Crucially, the fault labels in this dataset are dynamic rather than static, reflecting the time-varying nature of battery degradation. Users may observe that a specific cell's label evolves over consecutive cycles (e.g., from EI to ISC, or ASD to ACD). This is not a labeling error but a documentation of fault evolution—capturing the transition from a latent defect to a definitive failure. It should be noted that this study explicitly focuses on the nascent to mid-stage evolution of these faults. Data from the terminal thermal runaway stage is excluded to prioritize the development of early-warning algorithms capable of intervening before catastrophic failure. In conclusion, while we do not possess the OEM's proprietary diagnosis rule, the labels serve as a reliable validated by expensive and time-consuming post-mortem engineering analysis. They capture not only the distinct electrochemical signatures of each fault type but also their realistic dynamic evolution in field operations.

**Supplementary Note 2: Data processing**

Raw cloud data inevitably suffers from noise and corruption due to the stochastic nature of real-world collection and transmission. We implemented a robust preprocessing protocol to purify the dataset based on three distinct noise mechanisms:

1. Missing data: Instances of missing data (null entries) are treated as invalid samples and are rigorously filtered out to prevent algorithmic bias.
2. Abnormal data: Abnormal data manifesting as physically impossible values (out-of-range) are eliminated by applying domain-specific thresholds, ensuring all inputs adhere to valid operational ranges. For example, the range of voltage values is set to [0, 5], and the range of temperature values is set to [-50, 150].

**Supplementary Note 3: Analysis of complementarity and redundancy**

We constructed three-channel images inspired by the way visual deep learning models process RGB images [1]. Specifically, we designed Gramian angular field (GAF), Markov transition field (MTF), and recurrence plot (RP) representations, aiming for them to function in a complementary manner analogous to the three color channels in RGB images. This design provides the model with a richer and more diverse source of information for feature extraction. To evaluate the complementarity and redundancy among these representations, we analyzed their relationships using correlation and mutual information. Correlation quantifies the strength of linear dependence between two variables. In this study, we employed the Pearson correlation coefficient (PCC), which is mathematically defined as [2]:

$$r_{xy} = \frac{\sum_{i=1}^{n}(x_i - x)(y_i - y)}{\sqrt{\sum_{i=1}^{n}(x_i - \bar{x})^2} \cdot \sqrt{\sum_{i=1}^{n}(y_i - \bar{y})^2}}, \tag{1}$$

where $x_i$ and $y_i$ represent the $i$-th image matrix, respectively, and $\bar{x}$ and $\bar{y}$ are their respective means. The range of the correlation coefficient $r_{xy}$ is [-1, 1]. When the value is close to 1 or -1, it indicates a strong positive or negative correlation. And when it is close to 0, it suggests almost no linear relationship. On the other hand, we use the joint probability distribution to estimate mutual information (MI), which is used to measure the amount of information shared between different images. MI can capture nonlinear relationships, and its formula is defined as [3]:

$$I(x; y) = \sum_{x \in X} \sum_{y \in Y} p(x, y) \cdot \log \frac{p(x,y)}{p(x) \cdot p(y)}, \tag{2}$$

where $p(x, y)$ represents the joint probability distribution of $X$ and $Y$, while $p(x)$ and $p(y)$ represent the marginal probability distributions. The value of MI ranges from [0 to $+\infty$], with higher values indicating that the two variables share more information. Conversely, when the mutual information is 0, it means that the two are completely independent.

The analysis of images derived from different time-series transformations is presented in Supplementary Figure 4. It can be observed that the correlations among the three representations are generally low: the correlation coefficient between MTF–GAF is close to zero, MTF–RP exhibits a negative correlation, while GAF–RP shows almost no correlation. These findings indicate that the three methods do not exhibit significant linear redundancy in capturing temporal features, and in certain cases even reveal opposite trends. In contrast, the MI matrix uncovers another layer of relationships: the MI between MTF–GAF ranges from 0.54 to 1.02, MTF–RP from 0.48 to 0.72, and GAF–RP reaches the highest values, between 1.06 and 1.52. This suggests that although linear correlations are weak, the representations still share partial information, with GAF and RP showing particularly strong overlap in certain dimensions. Importantly, this shared information does not amount to complete redundancy but rather reflects their complementarity at different levels.

In summary, the combination of low correlation and non-zero mutual information provides strong evidence of complementarity among the three representations. Correlation analysis demonstrates their statistical independence, while mutual information analysis reveals that they still capture overlapping features in distinct representational spaces. This coexistence of independence and shared information enables the joint use of MTF, GAF, and RP to more comprehensively characterize the complexity of time-series images, thereby overcoming the limitations of single-feature representations and enhancing overall expressiveness. Furthermore, the results highlight that shorter data segments yield lower correlation and mutual information values, underscoring the effectiveness of multi-modal image representations in enriching information diversity in short-sequence scenarios.

**Supplementary Note 4: Metrics**

In this work, we employed multiple metrics to comprehensively evaluate the performance of the model, including accuracy, F1 score, area under curve (AUC) and average precision (AP) [4]. The values of these indicators usually fall within the range of 0 to 1. The closer they are to 1, the better the performance of the model. Accuracy is used to measure the number of samples that the model predicts correctly out of the total number of samples, and it is the most intuitive classification performance indicator. Its calculation formula is as follows:

$$\text{Accuracy} = \frac{TP+TN}{TP+TN+FP+FN}, \tag{3}$$

where $TP$, $TN$, $FP$ and $FN$ represent correctly predicted positive data, correctly predicted negative data, incorrectly predicted positive data and incorrectly predicted data. The F1 score is the harmonic mean of Precision and Recall, taking into account both the accuracy and coverage of the classification, and is suitable for scenarios with imbalanced categories, expressed as

$$\text{F1} = \frac{2\cdot\text{Precision}\cdot\text{Recall}}{\text{Precision}+\text{Recall}} = \frac{TP}{TP+\frac{1}{2}(FP+FN)}. \tag{4}$$

AUC represents the area under the receiver operating characteristic (ROC) curve (with the false positive rate $FPR$ on the x-axis and the true positive rate $TPR$ on the y-axis). The closer the AUC is to 1, the stronger the model's ability to distinguish between positive and negative samples. It can be expressed as

$$\text{AUC} = \int_0^1 TPR(FPR)\, d(FPR) = \int_0^1 \frac{TP}{TP+FN}\left(\frac{FP}{FP+TN}\right) d\left(\frac{FP}{FP+TN}\right). \tag{5}$$

AP is an indicator that quantifies the area under the Precision-Recall (PR) curve, representing the overall performance of the model at different recall rate thresholds and can be expressed as

$$\text{AP} = \int_0^1 \text{Precision}(\text{Recall})\, d(\text{Recall}) = \int_0^1 \frac{TP}{TP+FP}\left(\frac{TP}{TP+FN}\right) d\left(\frac{TP}{TP+FN}\right). \tag{6}$$

After calculating the macro average for each metric across all categories, it is presented as the overall evaluation metric. Furthermore, the training time, floating-point operations per second (FLOPs) and the number of parameters are used to evaluate the efficiency and computational cost of the model.

**Supplementary Note 5：Training details of methods**

**Method validation.** The validation and evaluation criteria for all methods are the same, and a total of 47968 samples from 99 EVs are obtained for the development and validation of the model. To ensure the fairness of the validation, we adopted the 5-fold strategy, and the ratio of training to testing set was 8:2 in each validation. Data originating from the same EV will not be present in both the training set and the testing set. During the training process, we set the batch size to 256 and select the AdamW algorithm as the optimizer. Moreover, we adopted a gradient learning rate, with an initial learning rate of 0.001 and early stopping mechanism to prevent model overfitting. All training and testing processes of the model are completed on PyTorch in a Python environment with a personal computer (Intel i7-14700K CPU, 32 GB RAM, and Nvidia RTX 4060 Ti).

**DeFault.** DeFault adopts a dual-stream architecture. A two-layer one-dimensional convolutional neural network (1D-CNN) specifically designed for sequential data and a two-layer two-dimensional convolutional neural network (2D-CNN) tailored for image data are employed to extract features from the respective modalities. Subsequently, a bidirectional cross-attention module is introduced to enable cross-modal learning and fusion of features from different sources. Finally, a multi-layer perception (MLP)-based classifier predicts the probability distribution of different fault types, with the class of highest probability output as the diagnostic result.

**Data fusion.** Data fusion integrates different modalities at the input stage, representing a simple yet flexible strategy. In this work, sequential data are first transformed into pseudo-images through a linear layer. These pseudo-images are then concatenated with the original image modality, and the fused representation is fed into a two-layer 2D-CNN for feature extraction. A classifier subsequently performs fault classification based on the extracted features.

**Feature fusion.** Feature fusion combines modalities after feature extraction, typically using a dual-stream structure. This model employs the same feature extraction and classification modules as DeFault. The key distinction lies in the fusion mechanism: instead of cross-attention, a fully connected neural network is used to integrate features across modalities.

**Decision fusion.** Decision fusion aggregates the independent outputs of each modality. In this model, the sequential and image modalities share the same feature extraction and classification modules as other multi-modal fusion approaches, but they operate independently. At the final stage, an additional classifier combines the outputs from both modalities to produce the aggregated diagnostic result.

**Variational Autoencoder (VAE).** VAE is a generative deep learning model capable of compressing and reconstructing features while preserving the underlying data distribution, thereby enabling unsupervised fault detection [5]. Once trained, the latent representations obtained from the VAE, together with labels, are used to train a classifier for multi-fault classification. Finally, different fault classes are predicted based on this supervised classifier. It is worth mentioning that the labels are only used for classifier training and the training of VAE is unsupervised.

**Linear Support Vector Classifier (LinearSVC).** LinearSVC is a classical supervised learning classifier that offers improved efficiency over traditional SVCs when handling large-scale datasets [6]. In this study, it serves as a baseline to evaluate the advantages of deep learning methods. LinearSVC is trained directly on either sequential or image modalities, with one classifier per fault category. During testing, all classifiers score the samples, and the category with the highest score is selected as the prediction.

**ResNet18.** ResNet18 is a widely used residual CNN for image classification [7]. By introducing skip connections between convolutional layers, ResNet18 ensures that original information is preserved in deep networks, thereby enhancing the capture of both fine-grained details and global features. In this work, we adopt a pre-trained ResNet18 model, modifying the final fully connected layer to output five classes, consistent with our multi-fault diagnosis task.

**Transformer.** Transformer architectures possess inherent advantages in processing sequential data [8]. Leveraging self-attention mechanisms, they effectively learn complex temporal dependencies within voltage signals. We use Transformers to encode and extract features from sequence data, and then use a fully connected classifier to predict the probability distribution between fault classes.

**Supplementary Note 6: Calculation of separation score**

To quantitatively evaluate the separability of different categories in the t-distributed stochastic neighbor embedding (t-SNE) embeddings, we designed and adopted a Separation Score. This metric jointly considers intra-class compactness and inter-class separability, thereby reflecting the discriminative effectiveness of categories in the low-dimensional embedding space. The Separation Score is defined as the ratio of inter-class distance to the sum of intra-class and inter-class distances, providing a comprehensive measure of class distribution separability. Specifically, for each category, we first compute the average Euclidean distance among all sample points within that category. The intra-class distance is then obtained by averaging these values across all categories. It is calculated as follows [9]:

$$D_{intra} = \frac{1}{C}\sum_{C=1}^{C}\frac{1}{N_C}\sum_{i,j\in c} d\left(x_i, x_j\right), \tag{7}$$

where $C$ is the number of classes, $N_C$ is the number of samples for class $C$, and $d(\cdot)$ is the Euclidean distance. The inter-class distance is then obtained by computing the centroid for each class and then the average distance between all the centroids, which can be expressed as follows [9]:

$$D_{inter} = \frac{1}{C(C-1)}\sum_{C_1\neq C_2} d(\mu_{C_1}, \mu_{C_2}), \tag{8}$$

where $\mu_c$ denotes the centroid of category. Finally, the separation score $S$ is defined as

$$S = \frac{D_{inter}}{D_{intra}+D_{inter}}. \tag{9}$$

When $S$ is close to 1, it means that the inter-class distance is much larger than the intra-class distance, and the class separation effect is good. On the contrary, when $S$ is close to 0, it means that the intra-class distance is large or the inter-class distance is small, and the class distribution overlaps.

**Supplementary Figure 1: Voltage curves of fault-free and faulty samples.** (A) Fault-free (FF). (B) Excessive inconsistency (EI). (C) Abnormal self-discharge (ASD). (D) Abnormal capacity degradation (ACD). (E) Internal short circuit (ISC). The three subfigures, from left to right, show voltage snapshots of three consecutive charging behaviors. Compared with FF cells, the voltage curves of different faulty cells have unique trajectories. It can be seen that both EI and ASC exhibit lower-than-normal voltage levels, but are difficult to distinguish clearly. The subtle difference between them is that the voltage deviation of EI cell and FF cells is always maintained at a certain level, while the voltage deviation of ASC cell will slightly increase when it is resting. ACD cell voltage is the highest cell voltage at the end of charging, but rapidly changes to the lowest cell voltage after charging. This is the performance of an abnormally lower capacity than FF cells. The voltage of the ISC cell is lower than it of FF cells regardless of the operating condition, and decreases continuously with the EV operation.

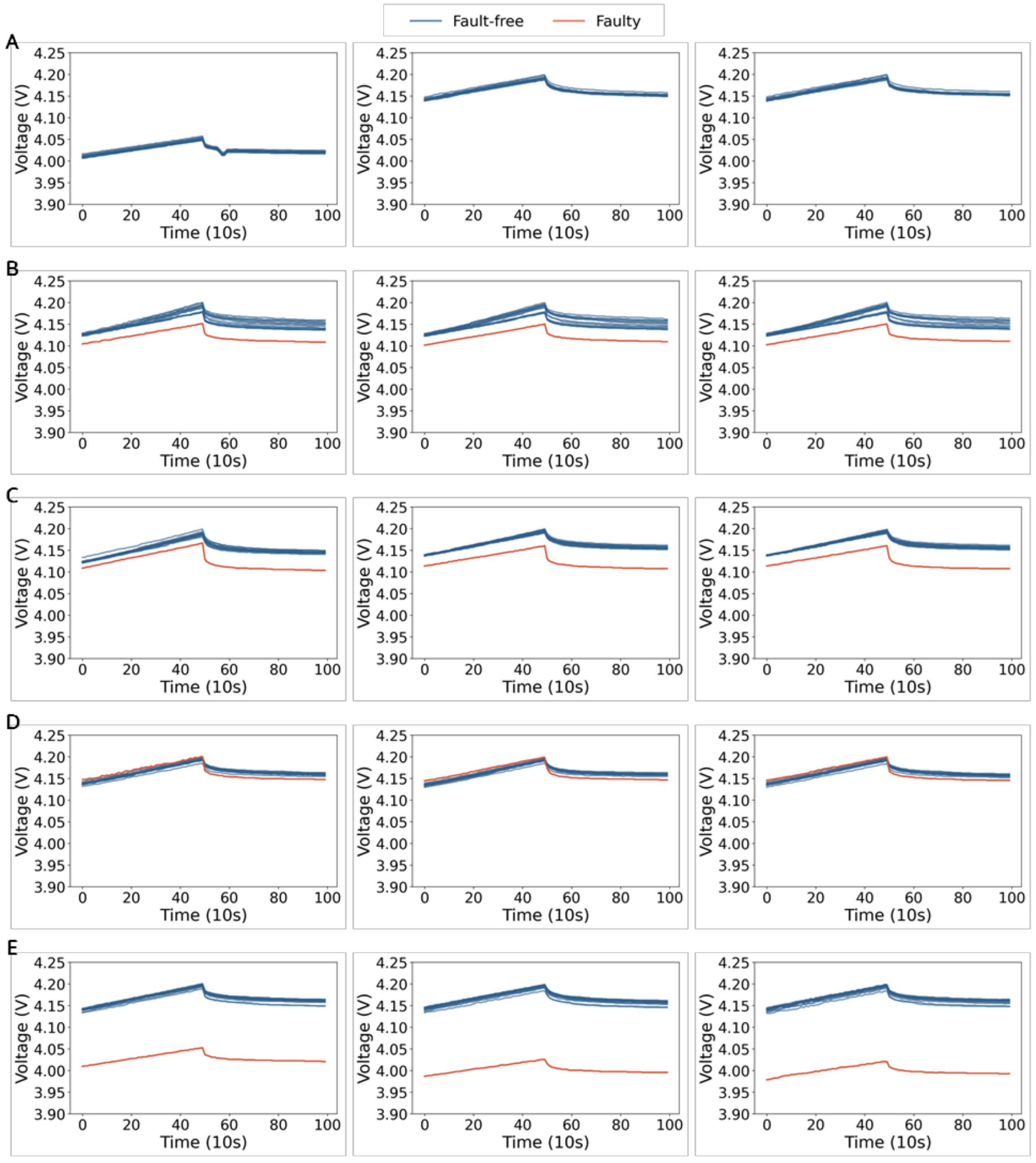

**Supplementary Figure 2: Concurrent battery faults in an EV.** (A) Cell voltage curves. (B) Voltage residuals subtracted from the median voltage. In the battery pack of this EV, there is an EI fault in cell 14, an ISC fault in cell 4 in the later stage. The voltage of cell1 deviates briefly and then returns to normal, which may be caused by the intense driving condition rather than the battery fault. Numerically, the voltage of the two faulty cells is always in the normal operating voltage range, and the maximum residual does not exceed 0.3V. Moreover, whether it is voltage or residual, there is a period of overlap between the ISC cell and the EI cell. This phenomenon indicates that although we can visually see the difference in the voltage trace, it is not distinguishable numerically.

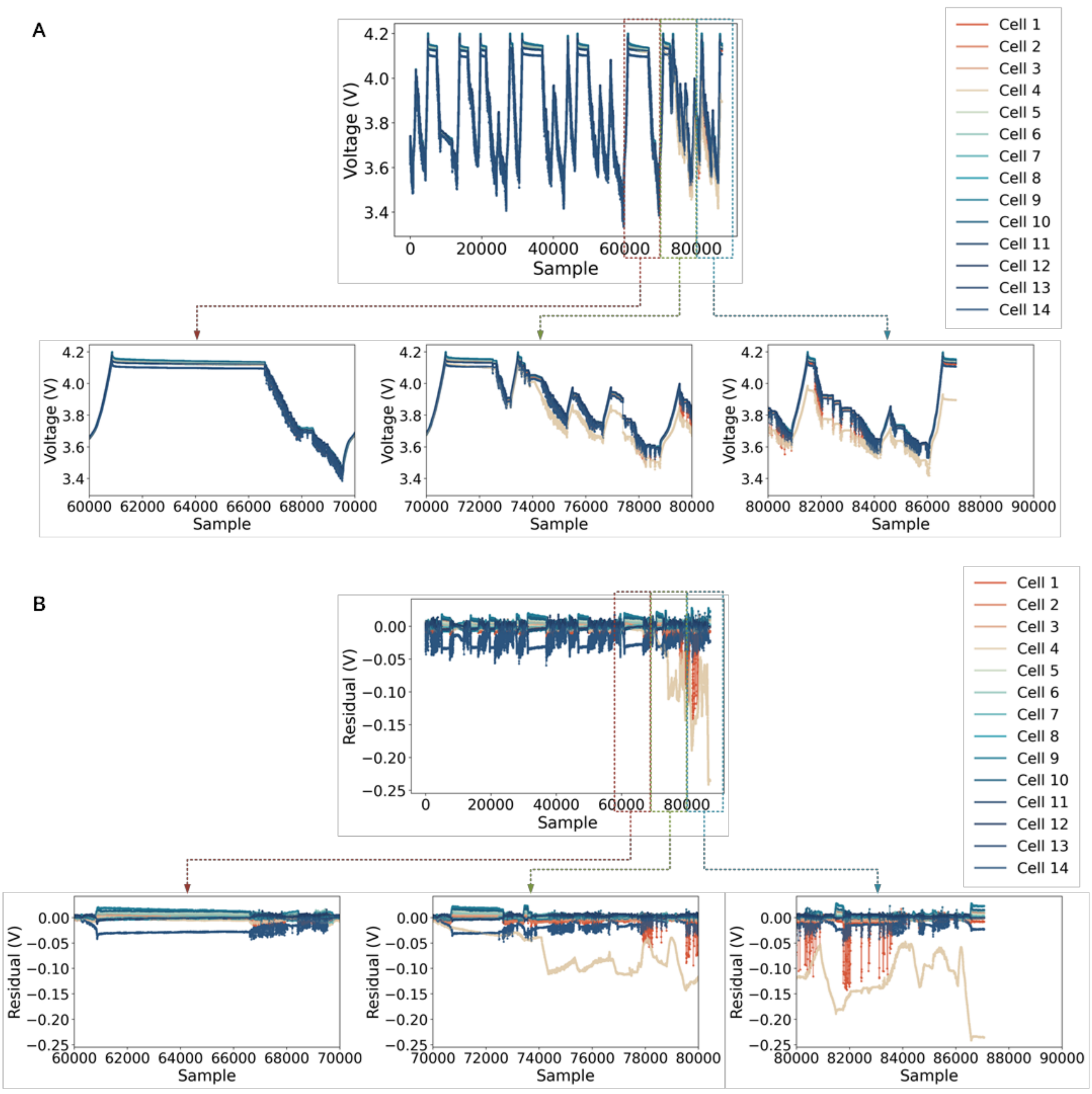

**Supplementary Figure 3: Extraction and presentation of multi-modal features.** From figures A-E, the raw voltage signals and extracted features of 5 labels (FF, EI, ASD, ACD and ISC) are presented. The top, middle, and bottom of each subfigure show the raw voltages, extracted sequence modalities, and extracted image modality, respectively. The raw voltages and sequence modalities are randomly selected from 100 cells of the same class, and the cells are in one-to-one correspondence according to color. The image modality is derived from a sequential modality above. Shown here is the average of three images, including GAF, MTF, and RP. Its color is used to highlight the difference of image modality between different labels.

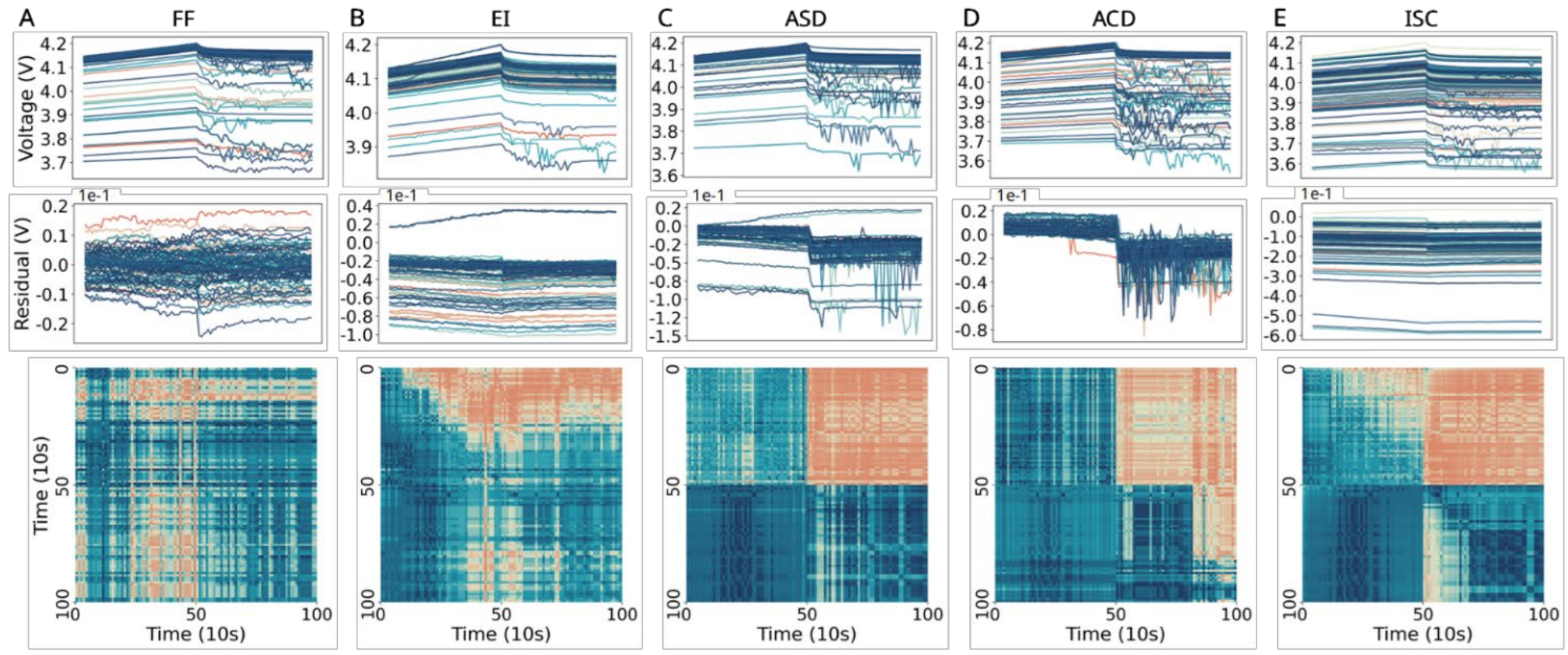

**Supplementary Figure 4: Analysis results of complementarity and redundancy.** The blue heatmap above shows the results of the correlation analysis, while the green heatmap below represents the mutual information between different images. (A) Based on data of 100 s. (B) Based on data of 200s. (C) Based on data of 500s. (D) Based on data of 1000s.

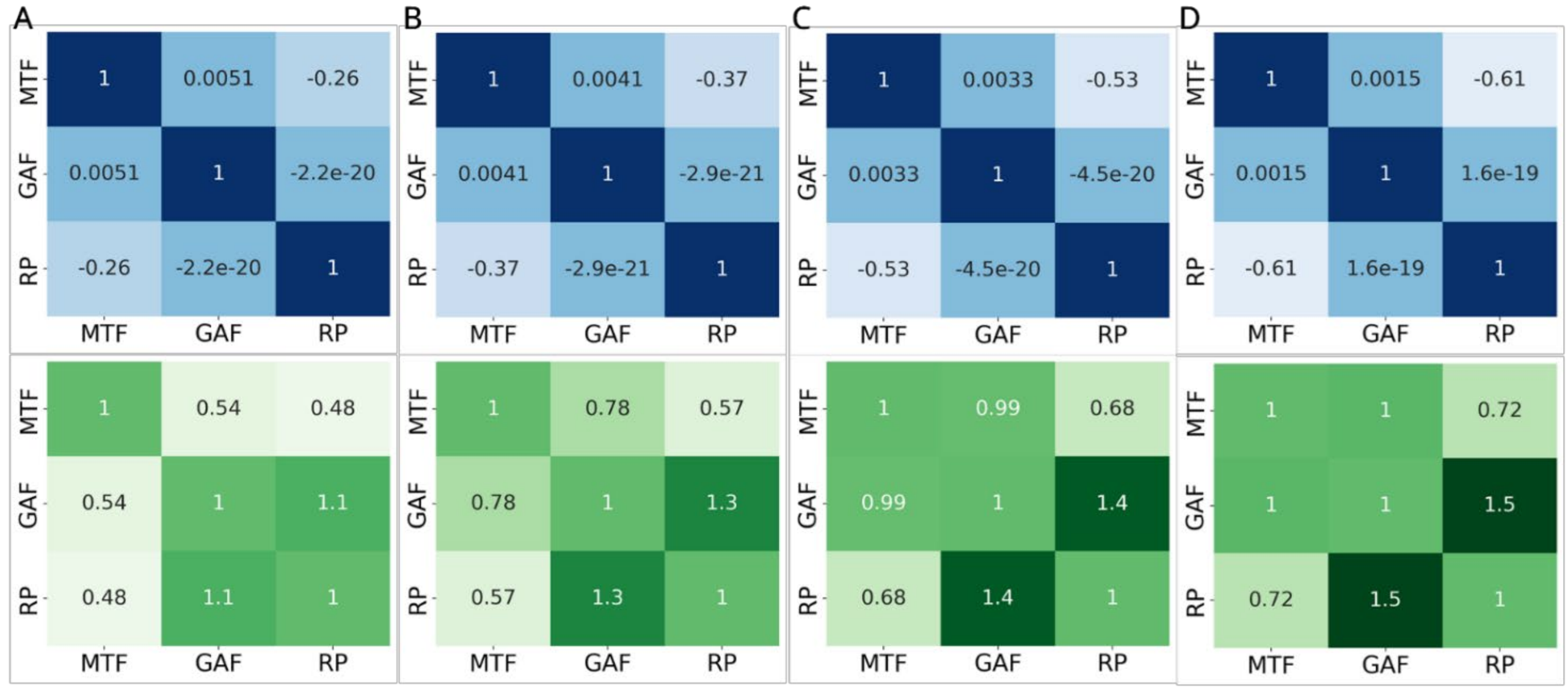

**Supplementary Figure 5: Fault detection performance analysis.** (A) Accuracy. (B) F1 score. In the fault detection task, the model achieved an average accuracy of 0.94–0.97 and an average F1 score ranging from 0.69 to 0.80. These results demonstrate that, despite the more sensitive evaluation criteria of detection tasks, the model consistently maintains stable and reliable performance in distinguishing between faulty and non-faulty samples. This highlights the robustness of DeFault under different task definitions and underscores its potential for practical deployment.

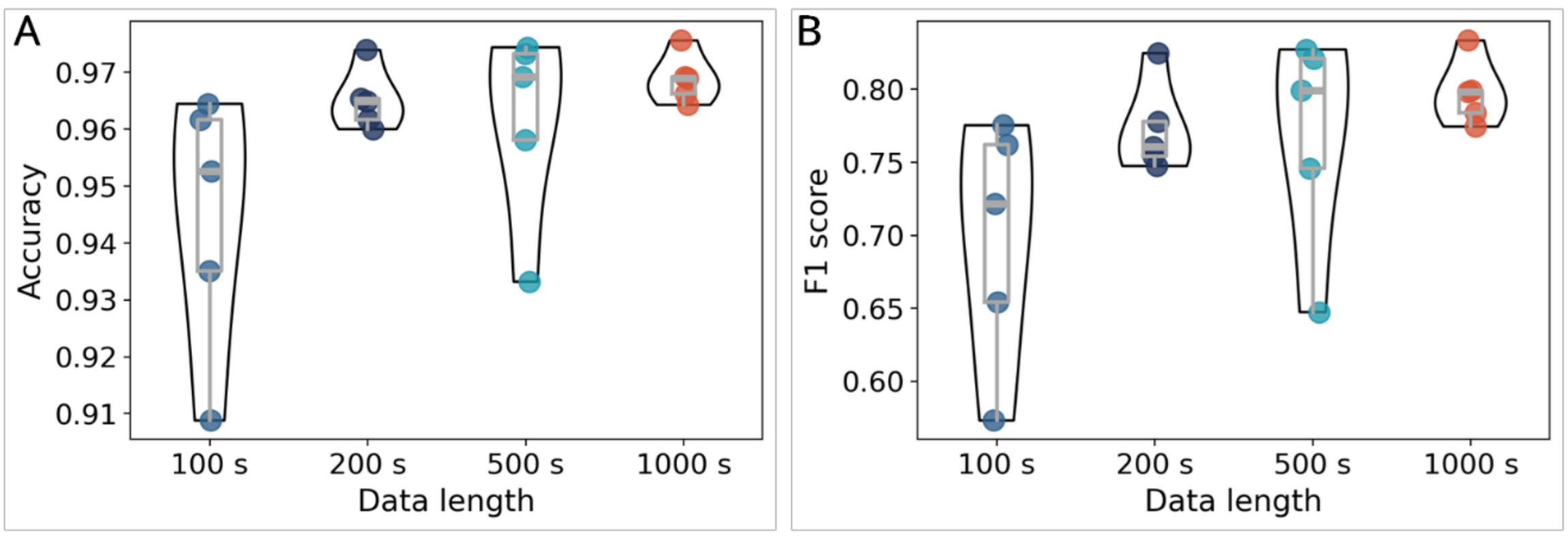

**Supplementary Figure 6: Performance comparison based on different data length.** In the comparative method, the markings I and S indicate that the input data are only image modality and sequence modality. Cross-modal representation the modal fusion structure of this model is the same as that of DeFault, the difference is that the feature input structures of its two-stream are either an image feature module or a sequence feature module. The left figure shows the average of the evaluation metrics, and the right figure shows the training time required by different models. (A) 100 s. (B) 200 s. (C) 1000 s.

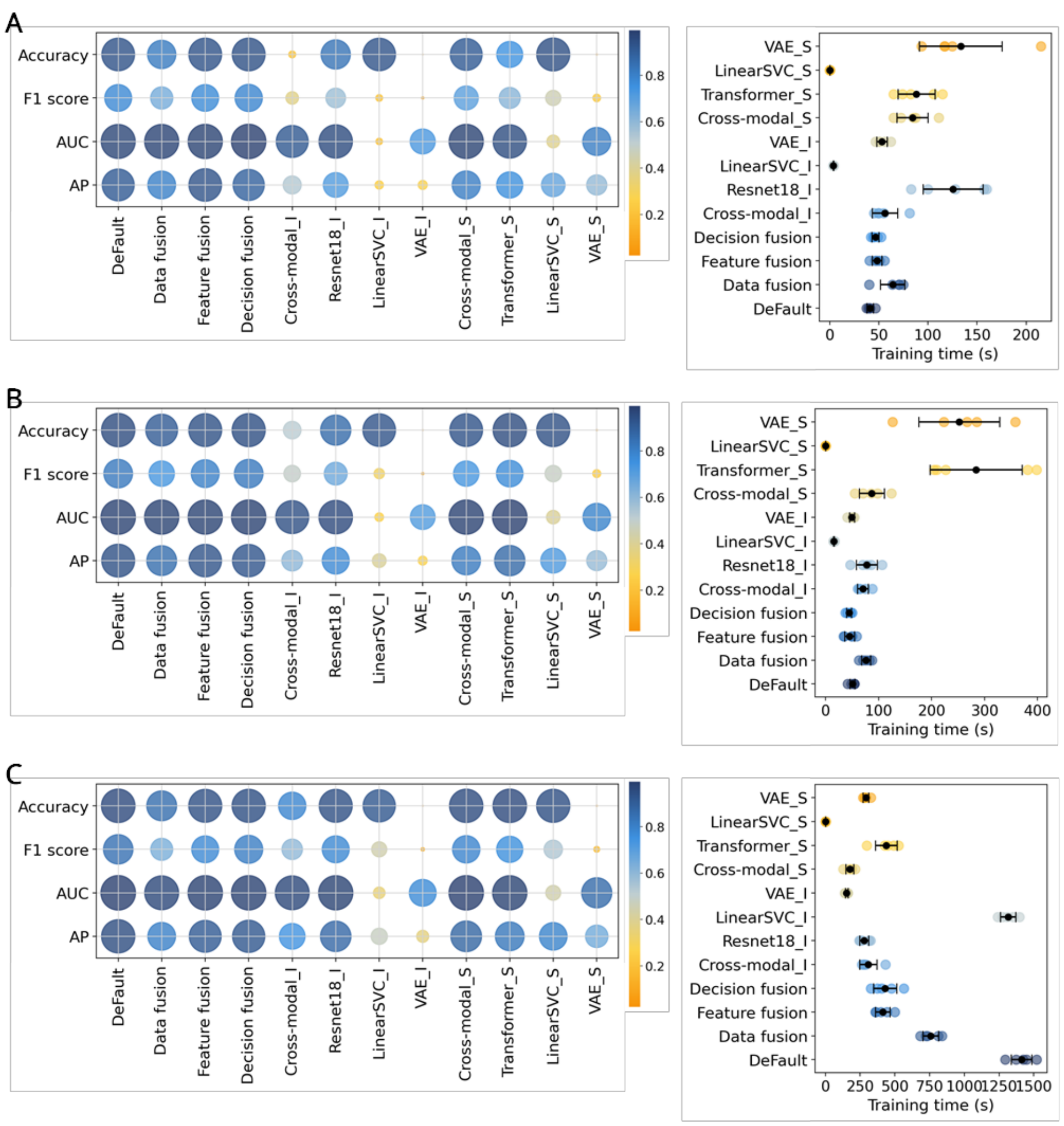

**Supplementary Figure 7: Visualization of the cross-attention diagnostic pipeline.** From top to bottom, the panel tracks the transformation from raw voltage snapshots to final diagnostic probabilities for five representative samples (FF, EI, ASD, ACD, ISC). *S* and *I* denote the Sequence and Image modalities, respectively. In the Cross-modal fusion row, the heatmaps visualize the bidirectional attention weights: "*I* to *S*" represents visual topology querying temporal precision, while "*S* to *I*" represents temporal trends seeking structural confirmation. Crucially, the attention "hotspots" (bright regions) shift dynamically across fault types, selectively activating mechanism-specific features (e.g., the depolarization in ACD or the voltage decay in ISC) to generate distinct latent variable barcodes for accurate classification.

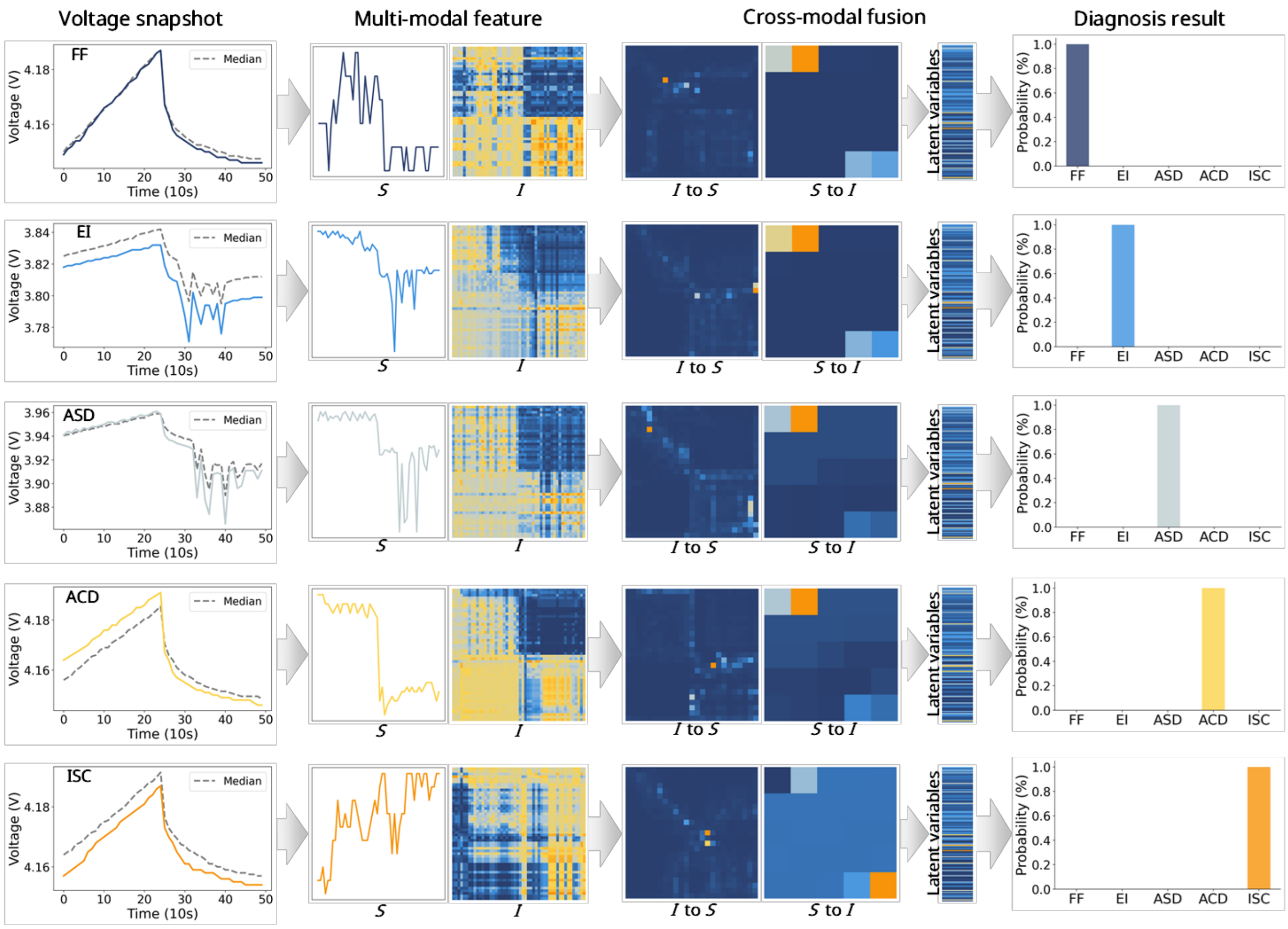

**Supplementary Table 1: Mechanism interpretation and feature manifestation of fault modes**

| Fault type | Mechanism | Behavior during state switch | Voltage manifestation |
|---|---|---|---|
| **FF** | Rapid unloading of double-layer capacitance and smooth dissipation of Li concentration gradients via Fickian diffusion. | Interface reaction rate drops sharply with current cutoff. Accumulated gradients relax gradually. | Minor transient voltage dip followed by smooth relaxation. Residuals remain near zero. |
| **EI** | Abnormal cells exhibit heterogeneity in particle size distribution, SEI thickness, reaction rate constant, diffusion coefficient, salt concentration, and porosity. | Interfacial recombination rate and gradient dissipation kinetics of abnormal cells differ from those of other cells. This results in asynchronous overpotential relaxation trajectories. | Voltage traces diverge in a fan-like manner. Residuals show multiple slopes and curvatures. |
| **ASD** | Elevated side reactions such as SEI growth, parasitic redox, and micro-leakage persist post-charging. | Even at near-zero external current, persistent interfacial current drives chemical potential shifts and continuous voltage decay. | Monotonic voltage decline superimposed on normal relaxation. Residuals show stable negative drift. |
| **ACD** | Loss of active lithium (LLI), active material (LAM), and interfacial degradation reduce effective capacity and transport. | Due to reduced capacity, SOC rises faster during charging, leading to steeper polarization buildup. | Pronounced voltage drop resembling a step-down and an extended relaxation tail. Residuals show a sharp negative jump with delayed recovery. |
| **ISC** | Low-resistance electronic bypass across electrodes induces continuous internal counter-reactions. | Internal current persists regardless of external load. Voltage recovery is suppressed and continuous decay is driven. | Voltage remains low across both phases. Residuals show strong negative bias and near-linear descent with low intra-class variance. |

**Supplementary References**